\documentclass[12pt]{ar}
\usepackage[english]{babel}
\usepackage{psfig,graphicx}

\onecolumn
\begin{document}

\title{New Spectroscopic Observations of the Post-AGB Star
            V354\,Lac\,=\,IRAS\,22272+5435}

\author{V.G.\,Klochkova, V.E.\,Panchuk, N.S.\,Tavolganskaya}

\date{\today}	     

\institute{Special Astrophysical Observatory RAS, Nizhnij Arkhyz,  369167 Russia}

\abstract{The strongest absorptions with the lower-level excitation
potentials $\chi_{\rm low}<1$\,eV are found to be split in the
high-resolution optical spectra of the post-AGB star V354\,Lac taken in
2007--2008 with the 6-m telescope of the Special Astrophysical Observatory.
Main atmospheric parameters, T$_{eff}$=5650\,K, $\log g$=0.2,
$\xi_t$=5.0\,km/s, and the abundances of 22 chemical elements in the
star's atmosphere are found. The excess of the $s$-process metals (Ba, La,
Ce, Nd) in the star's atmosphere is partly due to the splitting of strong
lines of the ions of these metals. The peculiarities of the spectrum in
the wavelength interval containing the LiI\,$\lambda$\,6707\,\AA{} line
can be naturally explained only by taking the overabundances of the CeII
and SmII heavy-metal ions into account. The best agreement with the
synthetic spectrum is achieved assuming $\varepsilon$(LiI)=2.0,
$\varepsilon$(CeII)=3.2, and $\varepsilon$(SmII)=2.7. The velocity field
both in the atmosphere and in the circumstellar envelope of V354\,Lac
remained stationary throughout the last 15 years of our observations.}

\titlerunning{\it Spectroscopy of the post-AGB star V354\,Lac}
\authorrunning{\it Klochkova et al.}
\maketitle

\section{Introduction}

The infrared source IRAS\,22272+5435 associated with the cool variable star
V354\,Lac\,=\,HD\,235858 (Sp\,=\,G5\,Iap [\cite{Hrivnak}]) is one of the
most interesting objects among protoplanetary nebula candidates (PPN).
This star stands out among related objects by its substantial photometric
variability: its B- and V-band magnitudes measured at two time instants
differed by 0\lefteqn{.}$^m$72 and 0\lefteqn{.}$^m$84,
respectively~[\cite{Hrivnak1991}]. The spectral energy distribution of
V354\,Lac has a two-peaked pattern typical of PPNs. Note that the total
energy emitted by the star at visual wavelengths is almost equal to that
emitted by the circumstellar envelope in the IR (see Fig.\,4 in[2]).

Objects observed during the short-lived PPN evolutionary stage are
intermediate-mass stars evolving away from the asymptotic giant branch
(AGB) toward the phase of a planetary nebula. The initial masses of these
stars are in the 3--8 ${\mathcal M}_{\odot}$ interval. A detailed
description of the evolution of intermediate-mass stars can be found,
e.g., in~[\cite{Block}]. We just recall the main moments of this process.
Having undergone the evolutionary stages of core hydrogen and helium
burning, these stars suffered an extensive mass loss in the form of
stellar wind (with mass-loss rates of up to $10^{-4}{\mathcal
M}_{\odot}$/yr). The loss of most of its mass leaves a star in the form a
degenerate carbon--oxygen core with a typical mass of about 0.6${\mathcal
M}_{\odot}$ and surrounded by the expanding gas-and-dust envelope. PPNs
are popular among the researchers first, because they allow us to study
the history of wind-driven mass loss and, second, because they offer a
unique opportunity to observe the result of stellar nucleosynthesis,
mixing, and dredge-up into the surface layers of the products of nuclear
reactions that took place during the previous stages of the star's
evolution.

The secular variability of the main parameters observed in some PPNs
stimulates spectroscopic monitoring of the most likely PPN candidates.
These observations revealed, e.g., the spectroscopic variability of the
optical components of the IRAS\,01005\,+\,7910~[\cite{IRAS01005}],
IRAS\,05040\,+\,4820~[\cite{IRAS05040}], and IRAS\,20572\,+\,4919~[\cite{V2324Cyg}]
infrared sources and the effective-temperature (T$_{eff}$) trend in
HD\,161796\,=\,IRAS\,17436\,+\,5003~[\cite{IRAS19475}]. Recall also more
than century-long observations of the parameters' evolution and chemical
composition of the famous highly evolved star FG\,Sge (see the review
by~[\cite{Jeffery}]). In this paper we report the results of
high-resolution spectroscopic observations of V354\,Lac made in 2007--2008
and compare the new data to those published earlier. The principal aim of
this paper is to reveal the possible spectral variability of the star,
study the velocity field in the star's atmosphere and envelope, and find
the fundamental parameters for the current epoch. Section\,\ref{observ}
briefly describes the methods of observations and data reduction;
Section\,\ref{results} analyzes the results obtained, and
Section~\,\ref{conclus} summarizes the conclusions.

\section{Observations and reductions of spectra}\label{observ}

We obtained our new spectroscopic data for V354\,Lac with the NES echelle
spectrograph mounted at the Nasmyth focus of the 6-m telescope of the Special
Astrophysical Observatory~[\cite{nes}].
Observations were performed using a  2048\,$\times$\,2048 CCD with an image
slicer~[\cite{nes}]. The spectral resolution was R$\ge$50000.
We took the first spectrum  (JD\,=\,2454170.58) in the  $4514-5940$\,\AA,
wavelength interval, and the next two spectra (JD\,=\,2454225.51 and
2454727.35), at longer wavelengths: \mbox{5215--6690} and
\mbox{5260--6760\,\AA} intervals, respectively. We use a modified ECHELLE context
[\cite{Yushkin}] of MIDAS package to extract one-dimensional vectors from
the two-dimensional echelle spectra. Cosmic-ray hits were removed via
median averaging of two spectra taken successively one after another.
Wavelength calibration was performed using a hollow-cathode Th-Ar lamp. We
determined the heliocentric radial velocities V$_{\odot}$ listed in
Table~\,\ref{data} by matching direct and mirrored images of observed line
profiles~[\cite{gala}]. To control and correct the instrumental offsets between the
spectra of the star and those of the hollow-cathode lamp, we use the H$_2$O and
O$_2$ telluric lines. Systematic errors do not exceed the measurement
errors (of about 1\,km/s based on a single line).

\begin{table*}[tbp]
\bigskip
\caption{Radial velocity V$_{\odot}$  at three observing times in 2007--2008.
The first line gives the radial velocity as measured from the spectrum taken
with the Lynx spectrograph (R=25000)~[\cite{lynx}] attached to the 6-m
telescope of the Special  Astrophysical Observatory of the Russian Academy of
Sciences. The number of spectral features measured to find the average
value is given  in parentheses. The last line gives the data from
Reddy et al.~[\cite{Reddy}.]}

\begin{tabular}{l|c|c|c|c|l|c}
\hline
JD=24...    &\multicolumn{6}{c}{V$_{\odot}$, km/s} \\
\cline{2-7}
            & metals & HI &\multicolumn{2}{c|}{D1,2\,NaI} & C$_2$ &DBs   \\
\cline{4-5}
            & & & blue & red & &     \\
\hline
48850.51    &$-$38.2  &$-$41.1 H$\alpha$&$-$50.2 &$-$13.2&$-$50.5(8) &  \\
54170.58    &$-$40.1  &$-$45.1 H$\beta$ &$-$50.6 &$-$14.4&$-$50.1(21)&   \\
54225.51    &$-$38.4  &$-$37.2 H$\alpha$&$-$51.1 &$-$13.4&     &$-$52:(5)\\
54727.35    &$-$38.0  &$-$34.6 H$\alpha$&$-$51.6 &$-$14.0&           &  \\
\hline
20.08.2000  &\multicolumn{6}{l}{$-$42.4 [\cite{Reddy}]} \\
\hline
\end{tabular}
\label{data}
\end{table*}

\section{Discussion of the results}\label{results}

\subsection{Peculiarity of the Spectrum}\label{peculiar}

\begin{figure}[tbp]
\includegraphics[angle=-90,width=0.9\textwidth,bb=32 30 570 790,clip]{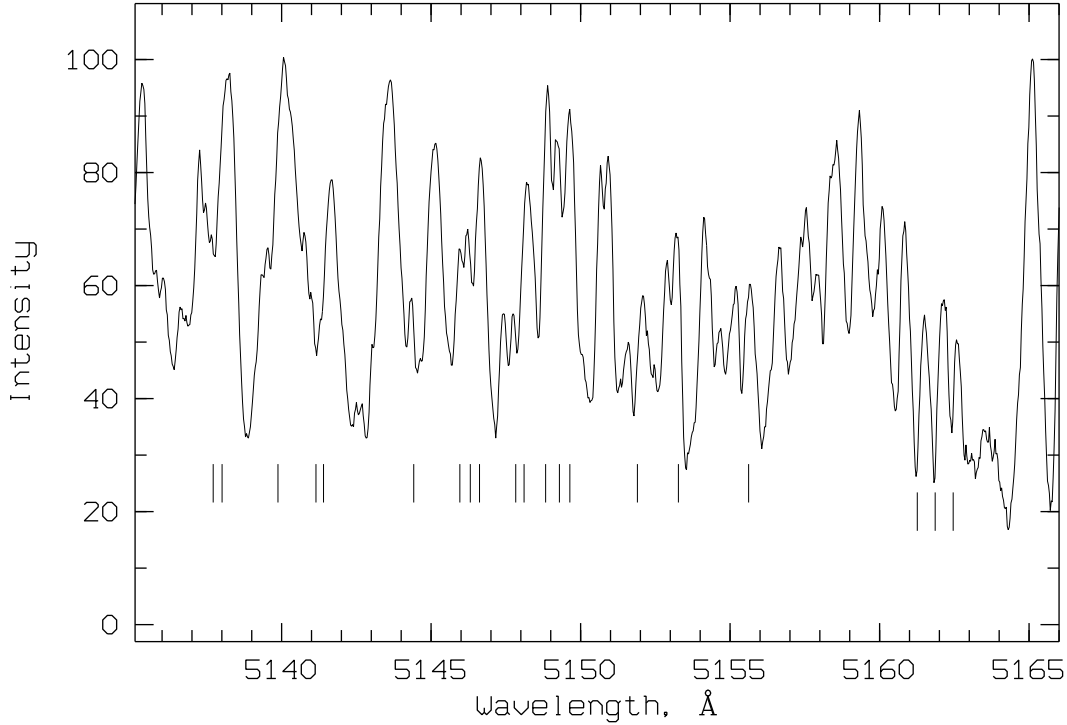}
\caption{The  C$_2$ (0;0) Swan band with the head at $\lambda$\,5165\,\AA{} in the
         spectrum of  V354\,Lac taken on JD=2454170.6. The vertical dashes indicate
         the lines of rotational transitions of this band used to find the expansion
         velocity of the envelope.}
\label{Swan}
\end{figure}

The main peculiarities of the optical spectrum of V354\,Lac were pointed out
by the authors of the very first low-resolution spectroscopic studies of this star.
Authors~\mbox{[\cite{Hrivnak, Hrivnak1991}]} found that, compared to the
spectrum of a normal supergiant of similar temperature, the H$\delta$ line
is weaker and BaII lines are stronger in the spectrum of  V354\,Lac. Latter,
in addition, also exhibits CN, C$_2$, and C$_3$ molecular bands. Most
of the above peculiarities also show up in our spectra taken in  2007--2008.
Figure~\,\ref{Swan} shows a portion of the spectrum with the Swan's
C$_2$ (0;0) oscillatory band with its head located at $\lambda$\,5165.2\,\AA{}.
Here we indicate the lines of the rotational transitions that we use to
find the expansion velocity of the envelope (see Section~\,3.2. below
for details). Low-excitation BaII lines are the strongest absorption features
in the spectrum of  V354\,Lac. The total equivalent width  W$_{\lambda}$ of each
BaII\,$\lambda$\,6141 and 6496\,\AA{} line exceeds 0.9\,\AA{}. The absorptions of
the ions of other  \mbox{$s$-process} elements (La, Ce) are equally strong, with
equivalent widths W$_{\lambda}>$0.3\,\AA{}.

\begin{figure}[tbp]
\includegraphics[angle=-90,width=0.9\textwidth,bb=40 80 570 790,clip]{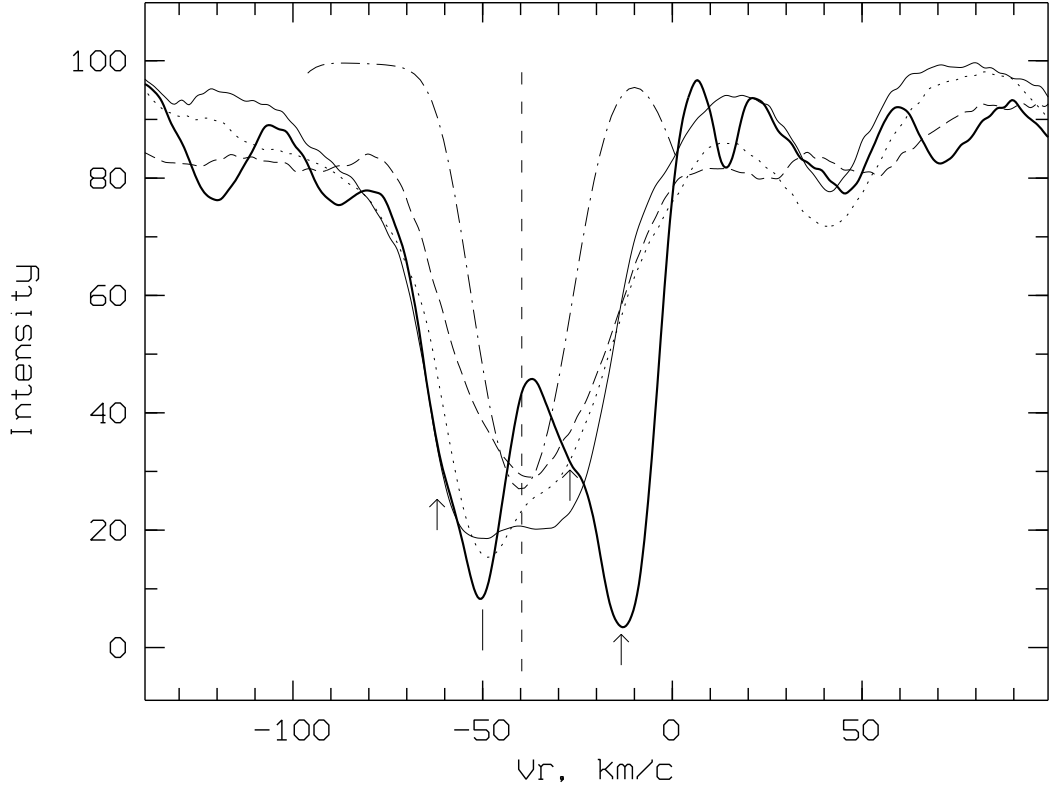}
\caption{Selected line profiles in the spectrum of V354\,Lac: the thick line shows
        the D1\,Na profile; the thin solid,  dashed-and-dotted, and dotted lines
        show the Ba\,II\,$\lambda$\,6141\,\AA{} profiles for three observing dates, and the
        dashed line shows the  H$\alpha$ profile. The vertical dashed line,  arrows, and
        vertical bar indicate the systemic velocity, interstellar components, and
        the circumstellar component of the  D1 line, respectively.}
\label{Prof}
\end{figure}

\begin{figure}[tbp]
\includegraphics[angle=-90,width=0.9\textwidth,bb=40 70 570 790,clip]{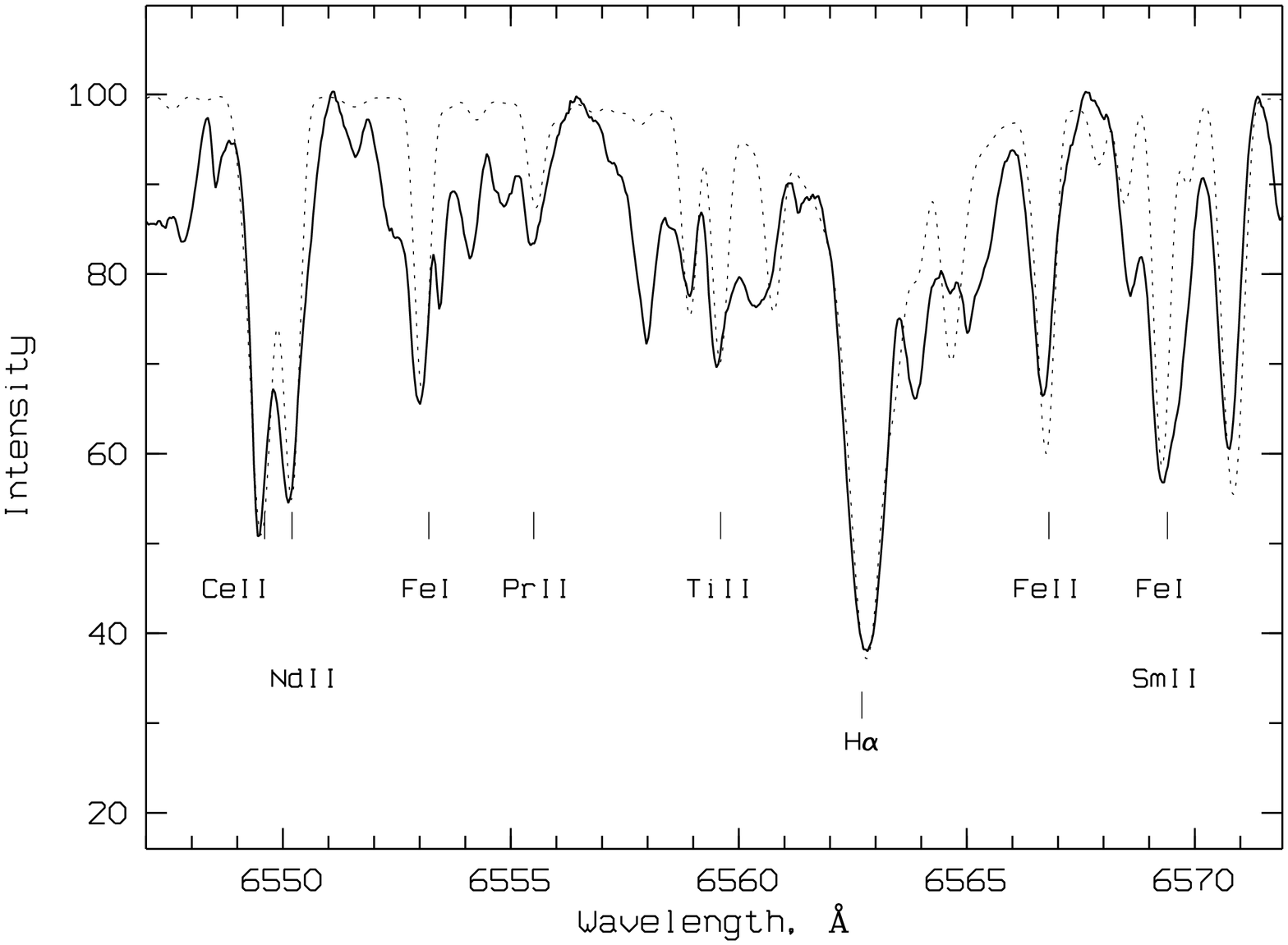}
\caption{A fragment of the spectrum of  V354\,Lac in the vicinity of the H$\alpha$
line. The dotted line shows the theoretical spectrum computed with
$T_{eff}$\,=5650\, K, $\log g$\,=\,0.2, and $\xi_t$\,=\,5.0\,km/s.
Abundances of the elements are adopted from Table~\,\ref{chemt}. Telluric
spectrum is not subtracted.}
\label{Synth_Halpha}
\end{figure}

The H$\alpha$ profile consists of a narrow-core absorption component with
wide wings (Fig.\,\ref{Prof} shows the profile of this line as observed on
JD=2454225.5). However, we found no decrease of the strength of this line
contrary to expectations based on~[\cite{Hrivnak1991,Hrivnak}]. As it is
evident from Fig.\,\ref{Synth_Halpha}, the observed H$\alpha$ profile in
the spectrum of V354\,Lac agrees well with the theoretical profile. This
fact suggests that the line must form in the atmosphere of the star with
negligible contribution from the envelope. Note that the cores of the
H$\alpha$ and H$\beta$ lines are shifted by \mbox{2--4\,km/s} with respect
to the velocity measured from photospheric lines of metals.

\begin{figure}[tbp]
\includegraphics[angle=-90,width=0.9\textwidth,bb=40 70 570 790,clip]{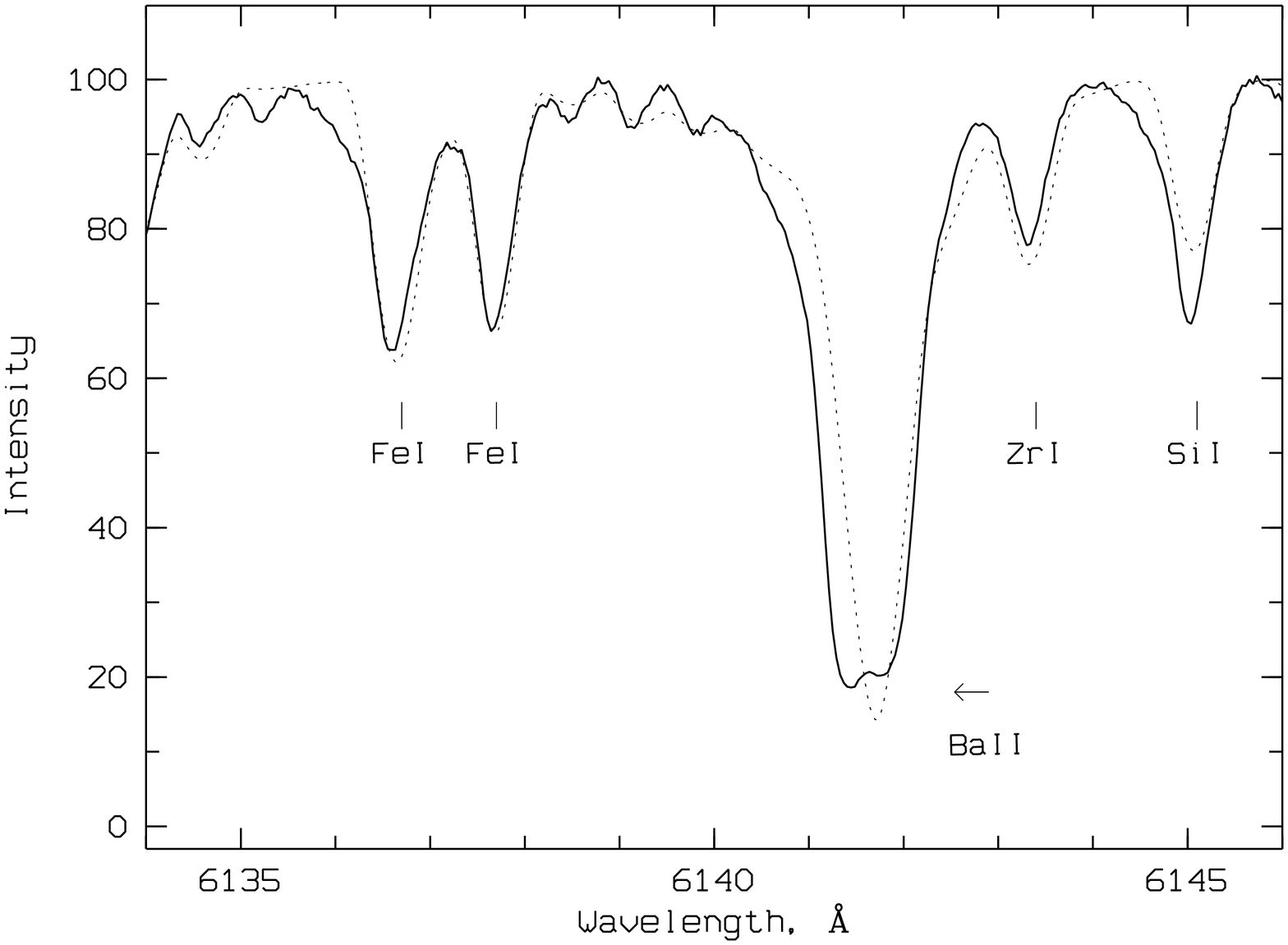}
\caption{A fragment of the spectrum of  V354\,Lac. The  FeI\,$\lambda$\,6136.7,
        FeI\,$\lambda$\,6137.7, BaII\,$\lambda$\,6141.7, ZrI\,$\lambda$\,6143.4,
        and SiI\,$\lambda$\,6145.1\,\AA lines are indicated. The dotted 
        line shows the theoretical spectrum computed with
        $T_{eff}$\,=5650\,K, $\log g$\,=\,0.2,
        $\xi_t$\,=\,5.0\,km/s, and barium abundance of $\varepsilon$(Ba)\,=\,3.94.}
\label{Ba6141}
\end{figure}

\begin{figure}[tbp]
\includegraphics[angle=-90,width=0.9\textwidth,bb=40 70 570 790,clip]{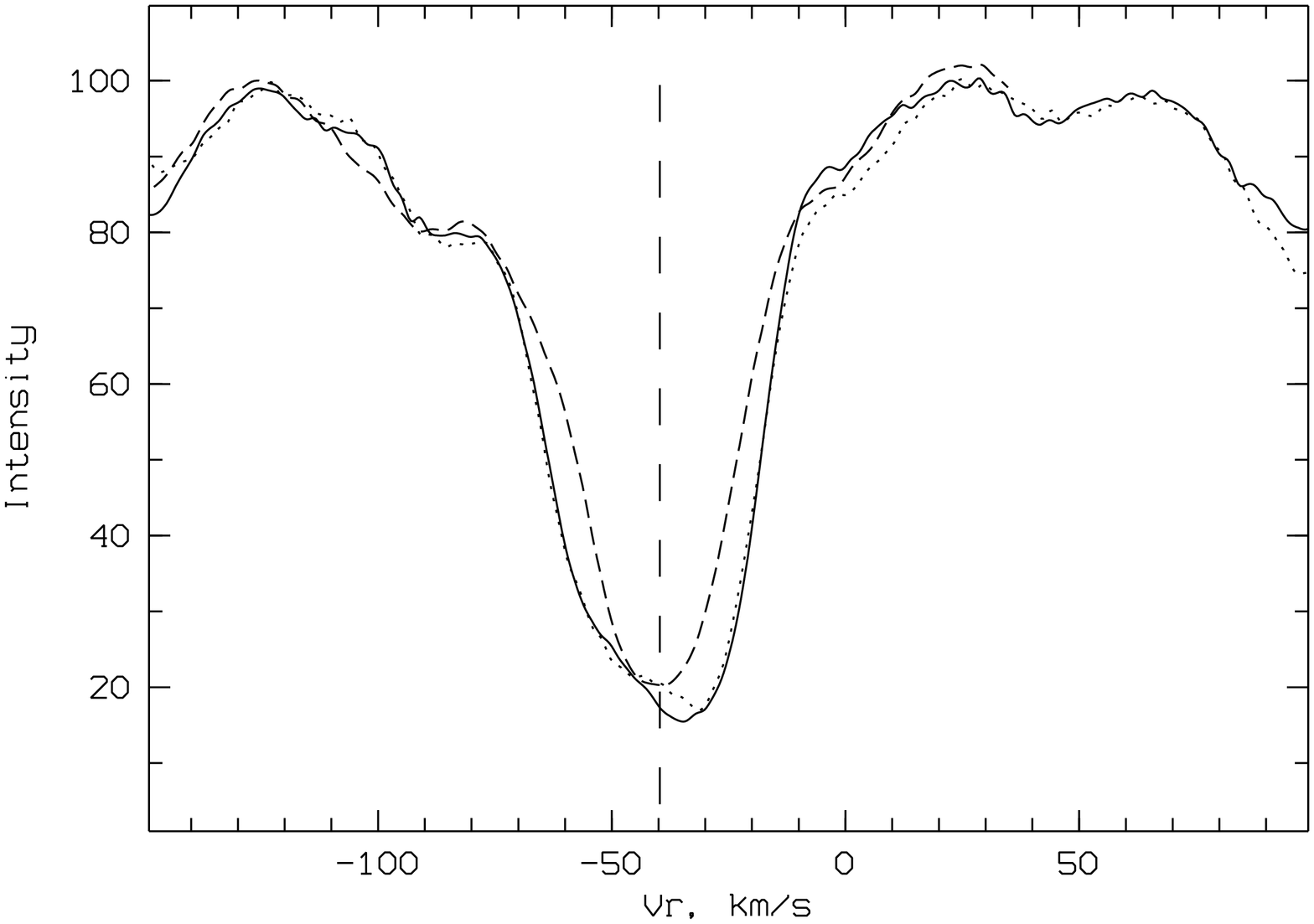}
\caption{The BaII\,5853\,\AA{} line profile in the spectra of  V354\,Lac
taken on different dates:  JD=2454170.6 (the solid line),
JD=2454225.5 (the dotted line), and JD=2454727.4 (the dashed line).
The vertical bar indicates the systemic velocity.}
\label{5853}
\end{figure}

The high spectral resolution of our instrument allowed us to reveal yet another hitherto unnoticed
feature of the optical spectrum of  V354\,Lac---the splitting of the
strongest absorptions of heavy-metal ions. This splitting is immediately
apparent in Fig.\,\ref{Ba6141} on the
BaII\,$\lambda$\,6141\,\AA~line profile with equivalent width
\mbox{W$_{\lambda}\approx$1\,\AA.} Such a splitting (or asymmetry of the profile
due to its flatter short-wavelength wing) is also observed in other
BaII~lines ($\lambda$\,5853 ¨ 6496\,\AA{}) and in the lines of
YII\,$\lambda$\,5402\,\AA{}, LaII\,$\lambda$\,6390\,\AA{}, and
NdII\,$\lambda$\,5234, 5293\,\AA{}. In particular, the asymmetry shows up
conspicuously in the profile of the  BaII\,$\lambda$\,5853\,\AA{} line
(Fig.\,\ref{5853}).

A comparison of our spectra taken on different dates with the same spectral resolution
suggests a slight difference between the absorption profiles. To illustrate the
variability of the profiles of strong lines, we show in Fig.\,\ref{5853}
a fragment of the spectrum with the  BaII\,5853\,\AA{} line as recorded on three
observing dates.

\begin{table}[tbp]
\caption{Abundances of the elements $\varepsilon(X)$ in the atmosphere of V354\,Lac.
Here $\sigma$ and n are the standard error and the number of lines used to compute
the abundances, respectively. The solar chemical composition is adopted
from~[\cite{Grev}].}
\medskip
\begin{tabular}{l| l | l | r | r| r|l}
\hline
\multicolumn{2}{c|}{The Sun} &\multicolumn{4}{c}{V354\,Lac}&    \\ [3pt]
\hline
X &$\varepsilon(X)$&$\varepsilon(X)$&$\sigma$&n&[X/Fe] & [X/Fe]$^1$ \\ [3pt]
\hline
LiI    & 3.31$^{2}$  & 2.00$^{3}$& &    &     &          \\
CI     & 8.39  &   9.08 & 0.22 &  7 & +1.40   &+0.98      \\
OI     & 8.66  &   8.98 & 0.15 &  4 & +1.03   &+0.50      \\
NaI    & 6.17  &   6.11 & 0.26 &  3 & +0.65   &+0.30      \\
MgI    & 7.53  &   7.23 & 0.01 &  3 & +0.41   &           \\
SiI    & 7.51  &   7.24 & 0.22 &  9 & +0.44   &+0.15      \\
SiII   &       &   7.26 &      &  2 & +0.46   &           \\
SI     & 7.11  &   7.17 & 0.25 &  3 & +0.77   &+015       \\
CaI    & 6.31  &   5.60 & 0.16 & 13 & +0.00   &+0.05      \\
ScII   & 3.05  &   2.93 & 0.16 &  8 & +0.59   &           \\
TiI    & 4.90  &   5.24 & 0.30 & 12 & +1.05   &+0.37      \\
CrI    & 5.64  &   5.13 & 0.25 &  7 & +0.20   &+0.27      \\
CrII   &       &   5.14 & 0.23 &  7 & +0.21   &           \\
MnI    & 5.39  &   4.46 & 0.18 &  7 &$-$0.22  &+0.22      \\
FeI    & 7.45  &   6.77 & 0.26 & 78 & +0.03   &0.00       \\
FeII   &       &   6.71 & 0.09 &  9 &$-$0.03  &0.00       \\
NiI    & 6.23  &   5.59 & 0.27 &  8 & +0.07   &           \\
NiII   &       &        &      &    &         &+0.02      \\
YII    &       &   3.50 &      &  2 & +2.00   &+1.81      \\
ZrI    & 2.59  &   3.64 & 0.44 &  3 & +1.76   &       \\
ZrII   &       &   3.32 &      &  1 & +1.44   &+1.31      \\
BaII   & 2.17  &   4.22 & 0.10 &  3 & +2.76   &       \\
LaII   & 1.13  &   3.32 & 0.31 &  6 & +2.90   &+2.27          \\
CeII   & 1.58  &   3.09 & 0.19 &  4 & +2.22   &+2.03          \\
PrII   & 0.71  &   2.48 &      &  1 & +2.48   &+1.71          \\
NdII   & 1.45  &   2.99 & 0.20 &  7 & +2.25   &+2.11          \\
EuII   & 0.52  &   1.40 & 0.28 &  3 & +1.59   &       \\
\hline
\multicolumn{7}{l}{} \\ [-8pt]
\multicolumn{7}{l}{\small 1---data adopted from~[\cite{Reddy}]}
\\[-3pt]
\multicolumn{7}{l}{\small 2---for meteorites~[\cite{Grev}]} \\[-3pt]
\multicolumn{7}{l}{\small 3---a result of spectral synthesis} \\[-3pt]
\end{tabular}
\label{chemt}
\end{table}

\subsection{Radial Velocities}\label{Velocity}

{\it Metal lines.} To find the average radial velocity V$_{\odot}$, we
measured the positions for a large set of absorption lines (about 300
features) minimally blended in the spectra of V354\,Lac. We selected the
lines based on a spectral atlas for the post-AGB star HD\,56126, which can
be viewed as a canonical post-AGB object~[\cite{Poland}]. The atlas was
prepared by Klochkova et al.~[\cite{Atlas}] based on echelle spectra taken
with the same NES spectrograph of the 6-m telescope. Compared to the
spectrum of HD\,56126, blending is stronger in that of V354\,Lac due to
its later spectral type (the effective temperature of HD\,56126 is
T$_{eff}$\,=\,7000\,K [\cite{Klochkova1995}]) and split and asymmetric
shape of many lines. Stronger blending and line asymmetry results in of
about 2\,km/s accuracy of a single-line V$_{\odot}$ measurement (standard
error $\sigma$) for our spectra taken in 2007--2008. Table~\,\ref{data}
lists the results of the V354\,Lac radial-velocity measurements based on a
set of spectral features. Column~2 gives the mean velocity
V$_{\odot}$(Met) inferred from metal-line measurements. The next columns
give the velocity measured from the H$\alpha$ and H$\beta$ lines of neutral
hydrogen, short- (blue) and long-wavelength (red) component of the D lines
of the NaI doublet, rotational lines of the Swan bands of the C$_2$
molecule, and spectral features identified with diffuse bands.
Table~\,\ref{data} also lists, besides the data based on the spectra taken
in 2007--2008, the results of our V$_{\odot}$ measurements made of the
spectrum taken earlier at the 6-m telescope with the Lynx echelle
spectrograph~[\cite{lynx}] with a resolution of R=25000. The last line
gives the average V$_{\odot}$ from~[\cite{Reddy}]. It follows from
long-term observations of Hrivnak~[\cite{Hrivnak2000}] that the amplitude
and period of radial-velocity variations in V354\,Lac are typical of a
PPN: radial velocity varies from --34 to \mbox{--41\,km/s} with a period
of $\rm 127^ d$. All mean velocities V$_{\odot}$ measured from metal
absorptions and listed in Table~\,\ref{data} are confined to a rather
narrow interval from --38 to --42\,km/s.

{\it Molecular spectrum.} The gas and dust envelope surrounding the central
star of the PPN shows up not only via IR excess and reddening, but also in the
features of optical spectra. Since molecular bands may form only in the
atmospheres of stars with the temperatures of  T$_{eff}\le$3000\,K, it is
evident that in case of a G5-type star they must form in the circumstellar
envelope. Our recorded spectra of V354\,Lac exhibit Swan C$_2$ molecular oscillatory bands.
The high spectral resolution of our observations allows the
positions of the rotational lines of the  (0;\,0) band to be measured with
rather high accuracy. We use the rotational-line wavelengths from the tables
published by Bakker et al.~[\cite{Bakker1997}] to measure the positions
of  21 rotational lines of the Swan band~(0;0) and determine the mean radial
velocity in the formation region of the band,
\mbox{V$_{\odot}(0;\,0)=-50.1\pm$0.2\,km/s.} Rotational lines of the Swan~(0;\,0)
band are easy to identify in the spectra because of their narrow profile compared
to photospheric lines (Fig.\,\ref{Swan}). Therefore the accuracy of
single-line position measurements for these lines is about 0.8\,km/s, which
is substantially better than that of the measurements based on photospheric
absorptions. The Swan band (1;\,0) in the short-wavelength part of the spectrum
\mbox{4712--4734\,\AA} is highly blended by photospheric lines, resulting in a
substantially lower accuracy, \mbox{V$_{\odot}(1;\,0)=-50.5\pm$1.0\,km/s.}

The offsets of the circumstellar features with respect to the systemic
velocity allow the expansion velocity to be found for the corresponding
parts of the envelope. The systemic velocity of IRAS\,22272+5435, V$_{\rm
lsr}=-27.5$\,km/s, was determined as the velocity of the center of the
\mbox{CO\,(1--0) emission profile~[\cite{Fong}],} the heliocentric systemic
velocity is V$_{\odot}^{\rm sys}\,=\, -39.7$\,km/s. Unlike the CO emission
lines, which form in the extended envelope expanding in all directions,
the observed absorption lines of molecular carbon form in the part of the
envelope that is located between the star and the observer and thus yield
the expansion velocity of the envelope (the formation region of the Swan
bands) to be V$_{\rm exp}$\,=\,10.8\,km/s. This result, which can be
viewed as the expansion velocity of the envelope measured from optical
spectra, agrees well with the expansion velocity of IRAS\,22272+5435,
V$_{\rm exp}$=10.6$\pm$1.1\,km/s, listed in the catalog of Loup et
al.~[\cite{Loup}], where numerous CO and HCN molecular-band observations of
circumstellar envelopes are collected. Note also that the expansion
velocity of the IRAS\,22272+5435 envelope is typical of those of the
circumstellar envelopes of post-AGB stars (see, e.g.,~[\cite{Loup}]).

The star's heliocentric velocity as measured from metal lines agrees well
with the systemic velocity V$_{\odot}^{\rm sys}\,=-39.7$\,km/s. This
agreement is indicative of the absence of a secondary component in the
system of IRAS\,22272+5435, or, more precisely, of the absence of a
stellar-mass secondary component. This is by no means a trivial result,
because chemical evolution, mixing, dredge-up of the products of nuclear
reactions into the surface layers of the star's atmosphere, mass outflow,
and the formation of the envelope may proceed differently in the presence
of a secondary companion.

By taking CO velocity maps for our Galaxy~[\cite{Vallee}], Galactic coordinates  ($l=103.3\degr$,
$b=-2.51\degr$),  and systemic velocity of IRAS\,22272+5435,  V$_{\rm lsr}=-$27.5\,km/s, into account,
we can conclude that the IR source is located between the Local and Perseus arms.

{\it Na\,I D-lines}. The lines of the NaI resonance doublet in the spectrum of V354\,Lac
have a complex structure. It follows from Table~\,\ref{data} and Fig~\,\ref{Prof},
which shows the D1 line profile, that the lines of the doublet contain two absorption components whose
positions correspond to the velocities of \mbox{V$_{\odot}=-$50} and $-$13\,km/s. It is evident that the line with 
\mbox{V$_{\odot}=-$50\,km/s} forms in the circumstellar envelope, where the envelope Swan bands of the
C$_2$ form. The second component with \mbox{V$_{\odot}\approx -$13\,km/s} \mbox{(V$_{\rm lsr}\approx -$27\,km/s)}
is of interstellar origin. The presence of this interstellar component with  \mbox{V$_{\rm
lsr}\approx -$27\,km/s} corroborates our hypothesis that V354\,Lac is located beyond the Local arm
our Galaxy.

It follows from Fig.\,\ref{Prof} that short-wavelength wings of the main
NaI absorptions may indicate the presence of components that are difficult
to resolve in our spectra: V$_{\odot} \approx -$57\,km/s (V$_{\rm
lsr}\approx -$70\,km/s), and V$_{\odot} \approx -$24\,km/s (V$_{\rm
lsr}\approx -$37\,km/s).

According to~[\cite{Georgelin}], the radial velocity is equal to
\mbox{V$_{\rm lsr}\approx -$10\,km/s} and \mbox{V$_{\rm lsr}\approx
-$55\,km/s} in the Local and Perseus spiral arms of our Galaxy,
respectively. Thus the distance to the Perseus arm,
d=3.6\,kpc~[\cite{Foster}], can be viewed as an upper estimate for the
distance to the source. Loup et al.~[\cite{Loup}] modeled the total flux
in IRAS\,22272+5435 and estimated the distance to the source to
be~d=2.35\,kpc.

{\it Diffuse interstellar bands.} V354\,Lac belongs to the subgroup of
PPNs, where circumstellar reddening is the main contributor to the color
excess~[\cite{Luna}]. This fact prompts the researchers to look for
diffuse (curcumstellar) bands (DBs) and diffuse interstellar bands~(DIBs)
in the spectrum of this star. Identification of DIBs and DBs in the
spectra of cool stars is a difficult task because these features are
blended with stellar lines. We looked for DBs in the long-wavelength
spectrum taken at JD\,=\,2454225.51. We used the wavelengths of the DIBs
adopted from the electronic supplement to the catalog~[\cite{Hobbs}] to
find in this spectrum of V354\,Lac several features that could be
identified with circumstellar bands (DBs). Unfortunately, we could not
measure the position of the band at $\lambda$\,6613.62\,\AA{}, which in
the spectrum of V354\,Lac is located in the wing of the strong YII
line\,$\lambda$\,6613.73\,\AA{}. The five most bona fide measured features
($\lambda$=5705, 5797, 6195, 6203, and 6269\,\AA{}) yield an average
velocity of \mbox{Vr(DBs)$_{\odot}\approx-52$\,km/s.} Note that this
average velocity measured from DBs agrees, within the quoted errors, with
the velocity of V$_{\odot} \approx-$50\,km/s measured from the
circumstellar component of the profile of NaI D lines. Such an agreement
may demonstrate the reality of the circumstellar analogs of DIBs. For a
more definitive conclusions higher-resolution spectra need to be taken in
a wider wavelength interval.

\subsection{Chemical Composition of the Atmosphere of  V354\,Lac}\label{chem}

\subsubsection{Parameters of the model}

To determine the main parameters of the stellar atmosphere---the effective temperature~$T_{eff}$
and surface gravity $\log g$, needed for the chemical composition and synthetic spectra to be calculated, we use the
grid of model atmospheres computed in terms of hydrostatic and LTE approximations for different
metallicities by Shulyak et al.~[\cite{Tsymbal}]. Fixing the main parameters---$T_{eff}$ and
$\log g$---is always a difficult task when computing the chemical composition of a star.
The problem complicates even further for objects with unclear evolutionary status and, hence, with
uncertain reddening, because in this case photometric data cannot be easily applied to find the
effective temperature. Moreover, the Balmer-line profiles in the spectra of PPNs may differ from
the corresponding profiles in the spectra of common supergiants~[\cite{BSAO}]. For these
reasons we determined the effective temperature of the star from the condition that FeI abundance
should be independent on the excitation potential $\chi_{\rm low}$ of the corresponding lines.
We selected surface gravity assuming ionization balance of iron atoms and ions
and microturbulence velocity $ \xi_t$, based on the condition that iron abundance should be independent
of the intensity of the line.

We already pointed out in Section~\,3.1 that  the observed H$\alpha$ profile in  V354\,Lac
agrees with the theoretical profile for the inferred parameters of the model, and this fact
is indicative of the viability of the model with the parameters considered. The reliability
of the model's choice is further corroborated by the fact that there's no dependence of the abundance on
excitation potential for the chemical elements with numerous lines in our spectra
(CaII, TiI, CrI, CrII). Moreover, a bona fide determination of microturbulence velocity results
in the fact that there's
no dependence of individual abundances on the equivalent widths of the lines used for the
computation. A typical accuracy of the model parameters  for
a star with an effective temperature of about 5500\,Š is, on the average,
\mbox{$\Delta T_{eff} \approx 100$\,K,} \mbox{$\Delta \log g
\approx 0.5\,dex $,} \mbox{$\Delta \xi_t \approx 1.0$ km/s.} An analysis
of the elemental-abundance errors due to the parameter errors mentioned
above and measurement errors of equivalent widths W$_{\lambda}$ leads us
to conclude that W$_{\lambda}$ errors are the main contributors to the
abundance uncertainties for the overwhelming majority of elements.

Standard test of the mutual consistency of the inferred parameters
consists in comparing the observed and synthetic spectra. We computed the
latter using SynthV code~[\cite{Tsymbal}]. A comparison of the spectra
shows them to agree satisfactorily with each other.
Figures~\,\ref{Synth_Halpha} and \ref{Ba6141} show, by way of an example,
fragments of the spectra in the vicinity of the H$\alpha$ and
BaII\,$\lambda$\,6141.7 lines. The computations were performed with
$T_{eff}$\,=5650\,K, $\log g$\,=\,0.2, $\xi_t$\,=\,5.0\,km/s, and the
elemental abundances from Table~\,\ref{chemt}.

We adopted the oscillator strengths $\log gf$ and other atomic constants
needed to compute the abundances of chemical elements from the VALD
database~\mbox{[\cite{VALD1,VALD2}]}. Table~\,\ref{chemt} lists the mean
elemental abundances relative to iron, [X/Fe], computed with the model
parameters $T_{eff}$\,=5650\,K, $\log g$\,=\,0.2, and
$\xi_t$\,=\,5.0\,km/s. We refer the elemental abundances in the star
studied to the solar chemical composition adopted from~[\cite{Grev}]. We
perform all chemical-composition computations using the programs written
by Shulyak et al.~[\cite{Tsymbal}] and adapted to PC/Linux platform. We
compute the plane-parallel models using the programs described by Shulyak
et al.~[\cite{Tsymbal}] with no corrections applied to allow for hyperfine
structure and isotopic shifts, which broaden NiI and BaII lines. The
scatter of the elemental abundances inferred from a set of lines is rather
small: the standard error ${\sigma}$ does not exceed 0.2\,dex in most of
the cases (see Table~\,\ref{chemt}), but increases for heavy nuclei. We
determine the main atmospheric parameters (T$_{eff}$, $\log g $, $\xi_t$)
from low-, and moderate-intensity lines with equivalent widths ${W \le
0.25}$\,\AA{}, because the plane-parallel and stationary atmosphere
approximation may be inadequate for describing more complex spectral
features.

\subsubsection{Chemical composition}

The V354\,Lac star was among the first candidate PPNs with a 21\,$\mu$m
feature in its IR spectrum whose atmospheres were found to exhibit large
overabundance of carbon and other $s$-process elements~[\cite{Zacs}].
Later, Reddy et al.~[\cite{Reddy}] used higher-resolution spectra to
perform a detailed analysis of the chemical composition and
radial-velocity pattern by the spectral features that form in the
atmosphere and circumstellar envelope of the star. Below we compare the
results obtained by Reddy et al.~[\cite{Reddy}] to compare with our results
since both of them have similar spectroscopic resolution. Let us now
analyze in more detail the elemental abudnances $\varepsilon$(X) by
grouping elements by the type of their synthesis.

{\it Light elements.} It follows from Table~\,\ref{chemt} that our carbon
and oxygen abundances are found reliably enough. We computed
$\varepsilon$(O) by two [OI] forbidden lines $\lambda$\,5577 and
6363\,\AA{}. The abundance inferred from the only allowed
OI$\lambda$\,6155\AA{} line available in our spectra agrees with that
inferred from forbidden lines. We found oxygen to be overabundant by
[O/Fe]=+1.03 and carbon to be highly overabundant, [C/Fe]=+1.40, and
C/O$>1$. We thus confirmed the status of V354\,Lac as a \mbox{C-rich}
star. This conclusion agrees with that of Reddy et al.~[\cite{Reddy}],
however, our carbon and oxygen overabundances are greater than those of
the above authors. Nitrogen lines were unavailable in the wavelength
interval recorded in our spectra.

{\it Lithium overabundance problem}. The absorption feature at
$\lambda$\,6707.8\,\AA{}, which is traditionally identified with a LiI
line, has a large equivalent width (W$_{\lambda}$\,=\,127\,m\AA{}) in the
spectrum of V354\,Lac, implying a very high lithium overabundance of
$\varepsilon$(LiI)\,=\,3.1. The problem of the high LiI overabundance
found by Za\~cs et al.~[\cite{Zacs}] and Reddy et al.~[\cite{Reddy}] in
case of V354\,Lac and by Reddy et al. for several other
PPNs~[\cite{Reddy1997,Reddy1999,Reddy}] prompts the researchers to
look for physical mechanisms of the production and fast dredge-up of
lithium isotopes into the atmosphere for evolved stars (see, e.g., a
review by, e.g., Lattanzio~[\cite{Lattanzio}]). Attempts to explain it by
contribution from $^6$Li isotope are unmaintainable due to the low
abundance of this isotope ([$^6$Li$/^7$Li]<0.1, see, e.g.,~[\cite{UFN}]).

\begin{figure}[tbp]
\includegraphics[angle=-90,width=0.9\textwidth,bb=40 70 570 790,clip]{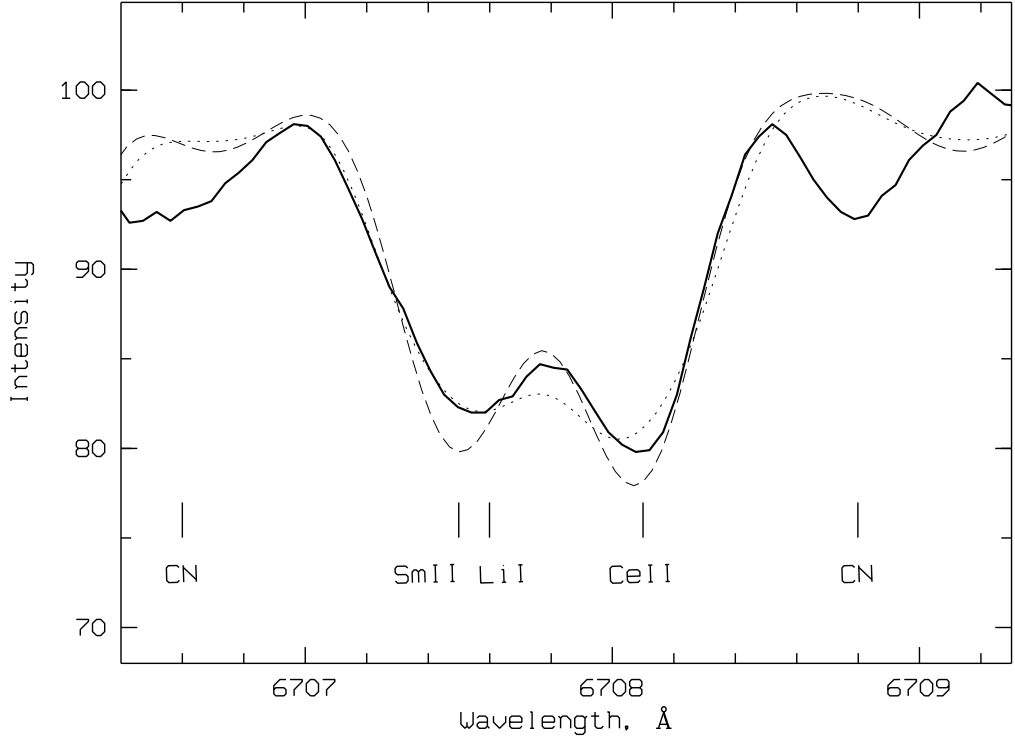}
\caption{Fragment of the spectrum of V354\,Lac in the vicinity of the LiI\,$\lambda$\,6707 feature
(the solid line). The synthetic spectrum is computed with  $T_{eff}$\,=5650\,K,
$\log g$\,=\,0.2, $\xi_t$\,=\,5.0\,km/s, and elemental abundances listed in Table~\,\ref{chemt}.
The dotted and dashed lines correspond to  $\varepsilon$(LiI)=2.0, $\varepsilon$(CeII)=3.2, $\varepsilon$(SmII)=2.7
and $\varepsilon$(LiI)=2.5, $\varepsilon$(CeII)=2.8, $\varepsilon$(SmII)=2.4, respectively. }
\label{Li6707}
\end{figure}

Recall that large overabundances of heavy metals and, in particular, of
samarium's ion were found in the atmosphere of V354\,Lac~[\cite{Reddy}].
Given that the LiI\,$\lambda$\,6707\,\AA{} line is blended with
SmII\,$\lambda$\,6707.5\,\AA{}, we considered it necessary to compute the
synthetic spectrum in order to refine the lithium abundance.

Besides the large equivalent width of the LiI line in the spectrum of
V354\,Lac, its interpretation poses yet another problem---that of
explaining the intense long-wavelength component near 6708.1\,\AA{} (see
Fig.\,\ref{Li6707}). Reddy et al.~[\cite{Reddy}] analyzed the possible
ways to interpret this component and favored the variant where it forms in
the circumstellar envelope. However, we consider this interpretation to be
unacceptable because it is inconsistent with the radial-velocity pattern
in the star's atmosphere and envelope, and propose a more natural
explanation based on CeII overabundance in the atmosphere of V354\,Lac.
Recently Reyniers et al.~[\cite{Reyniers}] showed that in case of high
cerium overabundance in the atmospheres of post-AGB stars the ions of this
elements (the CeII $\lambda$\,6708.099 line) contribute substantially to
the equivalent width of this feature. Our computations performed with
different contributions of heavy elements led us to conclude that good
agreement between observations and computations for the
$\lambda$\,6707--6708\,\AA{} range can be achieved by taking the
contribution of the SmII\,$\lambda$\,6707.47\,\AA{} and
CeII\,$\lambda$\,6708.099~lines~[\cite{VALD1,VALD2}] into account.

We found observations to agree best with the synthetic spectrum if the latter is computed with
$\varepsilon$(LiI)=2.0, $\varepsilon$(CeII)=3.2, and $\varepsilon$(SmII)=2.7
(this variant is shown by the dotted line in Fig.\,\ref{Li6707}). A somewhat
greater discrepancy with observations results if higher lithium abundance, $\varepsilon$(LiI)=2.5,
and lower cerium and samarium abundances, $\varepsilon$(CeII)=2.8, $\varepsilon$(SmII)=2.4 are
adopted (we show this variant by the dashed line in Fig.\,\ref{Li6707}). We thus
demonstrate that the peculiarities of the  $\lambda$\,6707--6708\,\AA{} fragment in the spectrum of V354\,Lac
can be naturally explained by correctly accounting for the overabundance of heavy elements observed in the
atmosphere of the star.

{\it Sodium abundance} is found from the moderate-intensity
NaI\,$\lambda$~5682, 6154, and 6160\,\AA{} lines with small NLTE
corrections~[\cite{Takeda1,Takeda2}]. Therefore the [Na/Fe]=+0.65 sodium
overabundance revealed may be mostly due to sodium synthesis during NeNa
cycle, which proceeds simultaneously with hydrogen burning in the CNO
cycle. The magnesium abundance as measured from the three MgI\,$\lambda$\,
5528, 5711, and 6319\,\AA{} lines, is also enhanced: [Mg/Fe]=+0.41. The
sodium-to-magnesium abundance ratio is [Na/Mg]=+0.24.

{\it Iron-peak elements.} The iron abundance in the atmosphere of
V354\,Lac, $\log\varepsilon$( FeI, FeII)=6.24, which is usually viewed as
the metallicity criterion, is lower than the solar iron abundance:
[Fe/H]=--0.76. Reddy et al.~[\cite{Reddy}] found a metallicity
\mbox{[Fe/H]=--0.81}, which agrees with our value within the errors of the
basic parameters findings. The abundances of chromium and nickel, which
belong to the iron group, also differ little from their normal values:
[CrI,CrII,NiI/Fe]=+0.16. The abundances of iron-group elements are, on the
whole, mutually consistent.

{\it Heavy metals.} Above, we already pointed that heavy elements
synthesized in the course of the $s$-process are overabundant in the
atmosphere of V354\,Lac~[\cite{Zacs,Reddy}]. It follows from
Table~\,\ref{chemt} that our computed abundances of Zr, Ba, La, Ce, Pr,
and Nd agree satisfactorily with the results of Reddy et al.~[\cite{Reddy}].
Furthermore, we also measured equally significant overabundance of
europium ([Eu/Fe]=+1.59), which is synthesized in the process of fast
neutronization, i.e., under the conditions of high neutron density.

The overabundance of heavy elements relative to iron was to be expected,
as it is often observed in the atmospheres of supergiants at the post-AGB
stage. The necessary physical conditions for efficient $s$-process and
subsequent dredge-up onto the surface of the matter enriched in heavy
nuclei are provided, in particular, in post-AGB stars (for the history and
current state of the problem see the review by Busso et
al.~[\cite{Busso}]). However, it is more common for the atmospheres of
these stars at the post-AGB stage to exhibit a deficit of $s$-process
elements~[\cite{BSAO,Winck}]. Among the PPN candidates about a dozen
objects were found to be overabundant in heavy metals, which are
synthesized via neutronization of iron nuclei at low neutron density (the
$s$-process). An analysis of the PPN spectra sample revealed that the
expected overabundances of $s$-process elements are observed in
carbon-enriched PPN atmospheres with the emission feature at 21\,$\mu$m in
the IR
spectra~[\cite{Klochkova1995,BSAO,Winckel,KSPV,Zacs,Reddy,IRAS20000}].
However, the overwhelming majority of PPNs exhibit neither carbon (the so
called O-rich stars) nor heavy-metal overabundance (see,
e.g.,~[\cite{Klochkova1995,IRAS19475,IRAS20056}]). Whether $s$-process
elements are or are not overabundant depends on the initial mass of the
star and on the mass-loss rate at the AGB stage---the factors that
determine the evolution of a particular star and the mass of the stellar
core. Modeling of the third mixing~[\cite{HerwigAustin}] shows that the
efficiency of the dredge-up of the reaction products increases with
increasing core mass (and hence with increasing initial mass) of the
post-AGB star. The correlation found between the overabundance of heavy
metals in the star's atmosphere and the 21\,$\mu$m feature of the IR
spectrum of the star's envelope need to be explained, and hence the sample
should be enlarged in order to study the most likely candidate PPN objects
in detail. Carbon overabundance combined with the presence of the
21\,$\mu$m feature leads us to conclude that the molecule that produces
this feature should contain carbon atoms~[\cite{Kwok1989}].

Recall, in connection with the observed overabundance of heavy elements,
that we found strong absorptions features of these elements in the
spectrum of V354\,Lac to have splitting core and/or asymmetric shape
(Fig.\,\ref{Ba6141}). In the spectrum of V354\,Lac the lines of these ions
(YII, ZrII, LaII, CeII, NdII and BaII) are enhanced to the extent that
their intensities become comparable to those of HI lines (cf.
Fig.\,\ref{Synth_Halpha} and \ref{Ba6141}). It is evident that
low-excitation lines that form in the upper layers of the stellar
atmosphere are influenced by the gaseous envelope. In case of
insufficiently high spectral resolution the intensity of envelope
components adds up to that of the components that form in the atmosphere.
To illustrate this, we show in Fig.\,\ref{Ba6141} a fragment of the
spectrum containing the BaII\,6141\,\AA{} line. It is evident from the
figure that the position of the short-wavelength component coincides with
that of the circumstellar NaD1 component (see Fig.\,\ref{Prof}). This
coincidence confirms that the complex profile of the BaII line contains
not only the photospheric component, but also features that form in the
circumstellar envelope. It follows from this that the heavy-nuclei
abundances found from the strongest absorption features in the spectrum of
V354\,Lac are overestimated, according to our estimates, by 0.2--0.4\,dex.
The abundances found from moderate-intensity lines are more realistic.

The ratio of the abundances of heavy (Ba, La, Ce, Pr, Nd) to those of
light (Y, Zr) $s$-process elements is known to be an important parameter
characterizing neutron exposure. The higher the neutron flux density, the
higher is the $hs/ls$ ratio. IRAS\,22272\,+\,5435 stands out among related
objects by its high heavy-to-light element ratio, $hs/ls=0.6$ (see, e.g.,
a compilation of these data in the paper by Reddy et al.~[\cite{Reddy}]).
Such a high $hs/ls$ ratio is typical for CH-stars~[\cite{Luck}]. The
$hs/ls$ ratio for IRAS\,22272\,+\,5435 is likely to somewhat decrease once
the discovered splitting of strong ion lines is taken into account.

{\it Separation of chemical elements in the envelope.} In stars with
gas-and-dust envelopes selective separation of chemical elements is known
to be a potentially efficient mechanism causing anomalous elemental
abundances in the atmosphere. In case of V354\,Lac we cannot rule out
completely condensation on dust grains. This conclusion is supported by a
slight relative overabundance of zinc and sulphur: [Zn/Fe]=+0.23 and
[S/Fe]=+0.15~[\cite{Reddy}]. The overabundance of zinc is within the
standard error for a single-line determination, whereas the sulphur
overabundance is reliably established based on six lines~[\cite{Reddy}].
Thus the iron deficit in the atmosphere (of about 0.2\,dex) can be partly
due to the condensation of iron atoms onto dust grains in the envelope of
the star.

\section{Conclusions}\label{conclus}

Core splitting or asymmetry (extended short-wavelength wing) is found in strong absorptions
with lower-level excitation potentials $\chi_{\rm low}<1$\,eV from the measurements of the
optical spectra of the post-AGB star V354\,Lac taken in 2007--2008 with the echelle
spectrograph of the 6-m telescope with a spectral resolution of R\,$\ge$\,50000. This applies
primarily to the strongest absorption features identified with the lines of heavy-element ions
(Ba, La, Ce, Nd).

The observed H$\alpha$ profile agrees well with the theoretical one computed with the
fundamental parameters of the star and normal (solar) hydrogen abundance. This agreement
suggests that the line must have formed in the star's photosphere and the envelope must have
contributed only slightly, and the photosphere has normal hydrogen abundance.

We found the main parameters of the star's atmosphere: T$_{eff}$=5650\,K, $\log g$=0.2,
$\xi_t$=5.0\,km/s, and the abundances of 22 chemical elements. The allowance for the
discovered splitting of the cores of strong absorptions may reduce the earlier found
overabundance of heavy elements. The peculiarities of the $\lambda$\,6707--6708\,\AA{}
spectral fragment can be naturally explained by taking the excess of heavy metals: the best
agreement between observations and synthetic spectrum was achieved for $\varepsilon$(LiI)=2.0,
$\varepsilon$(CeII)=3.2, and $\varepsilon$(SmII)=2.7

The radial velocity of the star as measured at three observational times in 2007--2008
agrees with earlier published data  within the quoted errors, suggesting no variations of the velocity
field in either the atmosphere or the circumstellar envelope of V354\,Lac over the last  15 years of
our observations.

\vspace{1cm}

{\it\bf Acknowledgments}. We are much grateful to Dr.~V.V.~Tsymbal for sharing Linux software for the
computation of stellar model atmospheres and synthetic spectra, and to M.V.~Yushkin for
assistance with observations.

This work was supported by the Russian Foundation for Basic Research (project
no.~08--02--00072\,a), the ``Extended objects in the Universe'' fundamental research program
of the Division of Physical Sciences of the Russian Academy of Sciences and the ``Origin and
evolution of stars and galaxies'' fundamental research program of the Presidium of the Russian
Academy of Sciences.

{}


\newpage
\begin{thebibliography}{99}

\bibitem{Hrivnak} 1. B.~J.~Hrivnak, Astrophys.~J, \textbf{438}, 341 (1995).

\bibitem{Hrivnak1991} 2. B.~J.~Hrivnak and S.~Kwok, Astrophys.~J., \textbf{371}, 631 (1991).

\bibitem{Block} 3. T.~Bl\"ocker, Astrophys. Space Sci. \textbf{275}, 1
            (2001).

\bibitem{IRAS01005}  4. V.~G.~Klochkova,  M.~V.~Yushkin, A.~S.~Miroshnichenko, et al.,
         V.~E.~Panchuk, K.~S.~Bjorkman.
        Astron. \& Astrophys., \textbf{392}, 143 (2002).

\bibitem{IRAS05040} 5. V.~G.~Klochkova, E.~L.~Chentsov, V.~E.~Panchuk, and
        M.~V.~Yushkin, IBVS \textbf{5584}, 1 (2004).

\bibitem{V2324Cyg}  6. V.~G.~Klochkova,  E.~L.~Chentsov, and  V.~E.~Panchuk,
                  Bull. Spec. Astrophys. Observ.,  \textbf{63}, 112 (2008).

\bibitem{IRAS19475}   7. V.~G.~Klochkova,  V.~E.~Panchuk, and N.~S.~Tavolzhanskaya,
                 Astron. Lett., \textbf{28}, 49 (2002).

\bibitem{Jeffery}  8. C.~S.~Jeffery and  D.~Sch\"onberner, Astron. \& Astrophys.,
                   \textbf{459}, 885    (2006).

\bibitem{nes} 9. V.~Panchuk, V.~Klochkova, M.~Yushkin and I.~D.~Najdenov,
 High resolution echelle spectrograph NES for visible and ground-based UV
regions. In: The UV Universe: stars from birth to death.
     in {\it Proceedings of the Joint Discussion No.\,4 during the IAU General
     Assembly of 2006} Ed. by A.~I.~Gomez de Castro and M.~A.~Barstow,
    (Editorial Complutense, Madrid, 2007), p.179.

\bibitem{Yushkin} 10. M.~V.~Yushkin, V.~G.~Klochkova, Preprint of the Special Astrophysical
                Observatory No. 206 (2005).

\bibitem{gala} 11. G.~A.~Galazutinov, Preprint of the Special Astrophysical Observatory
                  No.\,92 (1992).

\bibitem{Poland}  12. V.~G.~Klochkova,  E.~L.~Chentsov,  V.~E.~Panchuk, 
               N.~S.~Tavolganskaya, M.~V.~Yushkin.  Baltic  Astronomy, \textbf{16}, 155  (2007).

\bibitem{Atlas}  13. V.~G.~Klochkova,  E.~L.~Chentsov,  N.~S.~Tavolganskaya, and
         M.~V.~Shapovalov,  Bull. Spec. Astrophys. Observ.,
	 \textbf{62}, 162    (2007).

\bibitem{Klochkova1995} 14. V.~G.~Klochkova, Mon. Not. Roy. Astron. Soc.,  \textbf{272}, 710 (1995).

\bibitem{lynx} 15. V.~E.~Panchuk, V.~G.~Klochkova, I.~D.~Naidenov, et al., 
               Preprint of the Special Astrophysical Observatory No. 139 (1999).

\bibitem{Reddy} 16. B.~E.~Reddy, D.~L.~Lambert, G.~Gonzalez, and D.~Yong, Astrophys.~J.,
                \textbf{564}, 482 (2002).

\bibitem{Hrivnak2000} 17. B.~J.~Hrivnak and L.~Wenxian, in {\it Proceedings of the
       IAU Symp. No.\,177}, Ed. by R.~F.~Wing
       (Kluwer Acad. Publisher, Dordrecht, 2000), p.\,293.

\bibitem{Bakker1997} 18. E.~J.~Bakker, E.~F.~van~Dishoeck,  L.~B.~F.~M.~Waters, and
              T.~Schoenmaker, Astron. \& Astrophys., \textbf{323}, 469 (1997).

\bibitem{Fong}  19. D.~Fong, M.~Meixner, E.~C.~Sutton, et al.  A.~Zalucha, W.~J.~Welch.
              Astrophys.~J., \textbf{652}, 1626 (2006).

\bibitem{Loup}  20. C.~Loup, T.~Forveille, A.~Omont, and J.~F.~Paul,
              Astron. \& Astrophys. Suppl.   \textbf{99}, 291 (1993).

\bibitem{Vallee}   21. J.~P.~Vall\`ee. Asron.~J., \textbf{135}, 1301 (2008).

\bibitem{Georgelin} 22. Y.~P.~Georgelin and Y.~M.~Georgelin,
                        Astron. \& Astrophys., \textbf{6}, 349 (1970).

\bibitem{Foster} 23. T.~Foster and J.~MacWilliams, Astrophys.~J., \textbf{644}, 214 (2006).

\bibitem{Luna}  24. R.~Luna, N.~L.~J.~Cox, M.~A.~Sattore, et al.,
              Astron. \& Astrophys., \textbf{480}, 133 (2008).

\bibitem{Hobbs} 25. L.~M.~Hobbs, D.~G.~York, T.~P.~Snow, et al.
                Astrophys.~J., \textbf{680}, 1256  (2008).

\bibitem{Tsymbal}  26. D.~Shulyak,  V.~Tsymbal,  T.~Ryabchikova, et al.,  Ch.~St\"utz,  W.~W.~Weiss.
                    Astron. \&  Astrophys., \textbf{428}, 993 (2004).

\bibitem{BSAO}  27. V.~G.~Klochkova,  Bull. Spec. Astrophys. Observ.,
                        \textbf{44}, 5 (1997).

\bibitem{VALD1}  28. N.~E.~Piskunov, F.~Kupka, T.~A.~Ryabchikova, et al.,  W.~W.~Weiss, C.~S.~Jeffery.
                 Astron. \& Astrophys. Suppl.,  \textbf{112}, 525 (1995).

\bibitem{VALD2}  29. F.~Kupka,  N.~E.~Piskunov, T.~A.~Ryabchikova, et al., H.~C.~Stempels,  W.~W.~Weiss.
                Astron. \& Astrophys. Suppl.,   \textbf{138}, 119 (1999).

\bibitem{Grev} 30. M.~Asplund, N.~Grevesse, and A.~J.~Sauval, ASP Conf. Ser.,
               \textbf{336}, 25 (2005).

\bibitem{Zacs} 31. L.~Za\~cs, V.~G.~Klochkova, and V.E.~Panchuk, Mon. Not. Roy. Astron. Soc.,  
                  \textbf{275}, 764 (1995).

\bibitem{Reddy1997} 32. B.~E.~Reddy, M.~Parthasarathy, G.~Gonzalez, and E.~J.~Bakker,
               Astron. \& Astrophys., \textbf{328}, 331 (1997)

\bibitem{Reddy1999} 33. B.~E.~Reddy, E.~J.~Bakker, and B.~J.~Hrivnak,
                    Astrophys.~J., \textbf{524}, 831 (1999)

\bibitem{Lattanzio} 34. J.~Lattanzio,  in {\it Planetary Nebulae: Their Evolution
       and Role of Dregde-Up and Hot Bottom Burning}, Ed by. S.~Kwok, M.~Dopita, and R.~Sutherland.
       IAU Symp. \textbf{209}, 73 (2003).

\bibitem{UFN} 35. V.~G.~Klochkova, V.~E.~Panchuk,  Uspekhi Fiz. Nauk,
            \textbf{164}, 657 (1994).

\bibitem{Reyniers} 36. M.~Reyniers, H.~Van~Winckel, E.~Biemont, and P.~Quinet,
                Astron. \& Astrophys., \textbf{395}, L35 (2002).

\bibitem{Takeda1} 37. Y.~Takeda and M.~Takada-Hidai, Publ. Astro, Soc. Jap,  \textbf{46}, 395 (1994).

\bibitem{Takeda2} 38. Y.~Takeda, G.~Zhao, M.~Takada-Hidai, et al., Y.-Q.~Chen, Y.-J.~Saito, H.-W.~Zhang.
          Chin. J. Astron. Astrophys.,   \textbf{3}, 316 (2003).


\bibitem{Busso} 39. M.~Busso, R.~Gallino, and G.~J.~Wasserburg, Ann.Rev.
     Astron. \& Astrophys., \textbf{37}, 239 (1999).

\bibitem{Winck} 40. H.~Van Winckel, Astron. \& Astrophys.,  \textbf{319}, 561 (1997).

\bibitem{Winckel} 41. L.~Decin, H.~Van~Winckel, C.~Waelkens, and E.~J.~Bakker,
                  Astron. \& Astrophys., \textbf{332}, 928 (1998).

\bibitem{KSPV}  42. V.~G.~Klochkova,  R.~Szczerba,  V.E.~Panchuk, and K.~Volk,
                 Astron. \& Astrophys., \textbf{345}, 905 (1999).

\bibitem{IRAS20000} 43. V.~G.~Klochkova and T.~Kipper, Baltic Astronomy, 
                        \textbf{15}, 395 (2006).

\bibitem{IRAS20056} 44. V.~G.~Klochkova, V.~E.~Panchuk, E.~L.~Chentsov, and
                M.~V.~Yushkin,  Bull. Spec. Astrophys. Observ.,
		 \textbf{62}, 217 (2007).

\bibitem{HerwigAustin} 45. F.~Herwig and S.~M.~Austin, Astrophys.~J., \textbf{613}, L73 (2004).

\bibitem{Kwok1989}  46. S.~Kwok, K.~M.~Volk, and B.~J.~Hrivnak,
                    Astrophys.~J., \textbf{345}, L51 (1989).

\bibitem{Luck}  47. R.~E.~Luck and H.~E.~Bond, Astrophys.~J.~Suppl.,
                  \textbf{77}, 515 (1991).

\end{thebibliography}
\end{document}